\documentclass[preprint,11pt,authoryear]{elsarticle}
\usepackage[english]{babel}
\usepackage{tablefootnote}
\usepackage{multirow}
\usepackage{booktabs}
\usepackage{graphicx}
\usepackage{xspace}
\usepackage[caption=false]{subfig}
\usepackage{amsmath}
\usepackage{amssymb}
\numberwithin{equation}{section}

\newcommand{\Fermi}{\textit{Fermi}\xspace}
\newcommand{\fl}{\textit{Fermi}-LAT\xspace}

\begin{document}
\begin{frontmatter}
\title{Cut-off Characterisation of Energy Spectra of Bright Fermi Sources: Current instrument limits and future possibilities.}

\author[add1,add2]{C.~Romoli}
\ead{romolic@cp.dias.ie}
\author[add1]{A.~M.~Taylor}
\author[add1,add3,add4]{F.~Aharonian}

\address[add1]{Dublin Institute for Advanced Studies, 31 Fitzwilliam Place, Dublin 2, Ireland}
\address[add2]{Dublin City University, Glasnevin, Dublin 9, Ireland}
\address[add3]{Gran Sasso Science Institute, Viale F. Crispi 7, 67100 L'Aquila, Italy}
\address[add4]{Max-Planck-Institut f\"ur Kernphysik, P.O. Box 103980, 69029 Heidelberg, Germany}

\date{}

\begin{abstract}
In this paper some of the brightest GeV sources observed by the \fl were analysed, 
focusing on their spectral cut-off region. The sources chosen for this investigation 
were the brightest blazar flares of 3C~454.3 and 3C~279 and the Vela pulsar with a reanalysis
with the latest \fl software. For the study of the spectral cut-off we first explored the Vela pulsar spectrum,
whose statistics in the time interval of the 3FGL catalog allowed strong constraints
to be obtained on the parameters. We subsequently performed a new analysis of the flaring blazar SEDs.
For these sources we obtained constraints on the cut-off parameters under the 
assumption that their underlying spectral distribution is described by a power-law 
with a stretched exponential cut-off. We then highlighted the significant potential improvements
on such constraints by observations with next generation ground based Cherenkov telescopes, 
represented in our study by the Cherenkov Telescope Array (CTA). Adopting currently available 
simulations for this future observatory, we demonstrate the considerable improvement
in cut-off constraints achievable by observations with this new instrument when compared 
with that achievable by satellite observations.
\end{abstract}


\end{frontmatter}

\section{INTRODUCTION}

The gamma-ray emission from a broad range of both galactic and extragalactic objects has revealed
a multitude of effective particle accelerators. The gamma-ray energy spectrum of this emission can typically 
be described by a power-law distribution with a high energy cut-off, whose description we can generally 
encapsulate by a function of the form:
\begin{equation}
\frac{dN}{dE} = N_0\left(\frac{E}{E_{0}}\right)^{-{\Gamma}}\exp\left[-\left(\frac{E}{E_c}\right)^{\beta_{\gamma}}\right]\label{eq:plsec}\\
\end{equation}
where $E_0$ indicates the energy scale of the power-law region\footnote{In the fitting of the \fl data this parameter has been fixed to the value reported in the 3FGL catalogue \citep{3fgl_cat}.}; $\Gamma$ represents the power-law index of the particles, $E_c$ characterizes the position of the cut-off energy, while the parameter $\beta_{\gamma}$ determines the steepness of the cut-off (stretched for $\beta_{\gamma} < 1$, compressed for $\beta_{\gamma} > 1$). The determination of these parameters from the observational data is the focus of this work.

The importance of determining the shape of the cut-off region in the gamma-ray spectrum is directly connected with the cut-off region of the primary particles. A modified exponential cut-off for these parent particles naturally arises from the interplay between acceleration and energy loss rate. To avoid confusion we will call $\beta_{\gamma}$ and $\beta_{e}$ the cut-off parameters for photons and primary particles respectively.

When considering the acceleration of particles in the Bohm diffusion regime, for scenarios in which radiative losses can be safely neglected, we would naturally expect a simple exponential cut-off, with $\beta_{e}=1$. However, considering instead the acceleration of particles up to high energies for which radiative losses can no longer be ignored, the situation is more complicated. In the framework of diffusive shock acceleration \cite{zirak_2007} solved analytically the transport equation for electrons when dealing with Bohm  diffusion and synchrotron losses, obtaining $\beta_e=2$.

In the case of stochastic acceleration, already \citet{schlick_1985} and \citet{Ahar_ato_naha1986} demonstrated the formation of modified cut-offs in the particle spectrum when balancing acceleration and radiative losses. In this context, writing the momentum diffusion coefficient as $D(p) \propto p^{q}$, and the energy dependence of the time scale of the radiative losses as $\tau_{cool} \propto E^r$, the resulting cut-off of the primary particles can be described by $\beta_e=2-q-r$ \citep{stawarz_2008}. Typical values for the $q$ parameter are: $q=1$ for the Bohm case, $q=\frac{3}{2}$ and $q=\frac{5}{3}$ for, respectively, a Kraichnan or a Kolmogorov spectrum, and $q=2$ for the "hard-sphere" approximation.
Applied to the specific case of Bohm diffusion ($q=1$) and synchrotron losses ($r=-1$), we obtain a $\beta_e=2$. Thus, once again, the inclusion of synchrotron cooling in the acceleration process can lead to a sharpening of the cut-off shape.

The effect of $\beta_e$ is to modify the cut-off of the primary particles, and consequently the resultant cut-off in the photon spectrum emitted. This emitted spectrum may itself be described by a stretched cut-off, with stretching parameter $\beta_{\gamma}$. For the specific case of synchrotron emission, this parameter $\beta_{\gamma}$, relates to the parent population parameter through the relation $\beta_{\gamma}=\frac{\beta_e}{\beta_e+2}$ \citep{fritz_1989}, indicating that a cut-off in the photon spectrum with a compressed exponential shape is incompatible with a synchrotron origin and acceleration taking place in the Bohm regime where we would expect $\beta_{\gamma}=0.5$.

When we deal instead with inverse Compton processes, the emitted spectrum of the scattered photons is affected by both the electron distribution and the target photon field. The outcome is also affected by the cross section of the interaction, with the resultant spectrum depending on whether the process occurred: in the Thomson regime ($\varepsilon_e\varepsilon_{\gamma}^{bg} \ll \left(m_e c^2\right)^2$) or Klein-Nishina ($\varepsilon_e\varepsilon_{\gamma}^{bg} \gtrsim \left(m_e c^2\right)^2$) regime. Analysis of the various processes has been carried out by \citet{lefa_cutoff2012} taking into account different photon fields. They showed that in the Klein-Nishina regime, due to the fact that the electron loses almost all of its energy in a single interaction with the photon, the spectrum of the latter resembles the spectrum of the parent electrons, with $\beta_{\gamma}=\beta_{e}$. In the Thomson regime, instead, the photon spectrum is always stretched with $\beta_{\gamma} < \beta_e$. For example in the case of Inverse Compton on a Planckian photon seed field $\beta_{\gamma}=\frac{\beta_e}{\beta_e+2}$, while when considering a Synchrotron Self Compton mechanism, the gamma-ray photon spectrum will have a cut-off described by $\beta_{\gamma}=\frac{\beta_e}{\beta_e+4}$.

Another important channel for the production of gamma rays is proton-proton interactions where the gamma rays are emitted through the production and decay of secondary neutral mesons (mainly $\pi$ and $\eta$). Once the description for emissivity of the $\pi^{0}$-meson is taken into account, it is possible to show also here that a stretching of the cut-off in the photon spectrum also occurs \citep{2006PhRvD..74c4018K,2014PhRvD..90l3014K}.

Objects for which this cut-off sits in the GeV domain, presently may be most effectively probed by the Large Area Telescope (LAT) onboard the \Fermi\ satellite. This is a pair conversion telescope capable of reconstructing the direction of incoming photons with energies between 20~MeV and more than 300~GeV \citep{atwood_2009}. Unfortunately the measurement of the spectrum in the cutoff regime requires large photon statistics and this is available only for a limited number of Fermi sources.

In section 2 a subset of some of the brightest objects observed by the \fl is considered. This set of objects contains the Vela pulsar and 2 bright flaring AGNs. Utilising \Fermi data, the spectra of these bright objects with the highest statistics in the GeV range are used to constrain the photon spectral shape in the cut-off region as a tool for probing the acceleration, escape, and radiative loss processes giving rise to the particle energy distribution in this region. The list of these objects is provided in table \ref{tab:ana_sources} along with the time window for which we extracted the spectrum. In the following subsections we report the results of this analysis. In section 3, the potential improvements brought about by next generation instruments are considered. The benefits from the increase of the collection area on the data quality are demonstrated to be considerable. In section 4, our conclusions on the present and future ability to accurately determine the underlying particle cut-off shape using gamma-ray instruments are made.

\begin{table}[ht]
\centering
\footnotesize
\caption{Sources and type of event analysed. In the last column is reported the MJD interval from which the SED has been extracted.}
\footnotesize
\begin{tabular}{l|c|c|c|c}
\toprule
Object      & Class & Event type & Analysed period  & MJD interval\\
\midrule
\midrule
3C 454.3    & AGN (FSRQ)    & Flare             & Nov. 2010 & 55516 - 55523 \\
3C 279      & AGN (FSRQ)    & Flare             & June 2015 & 57187 - 57190 \\
Vela PSR\tablefootnote{3FGL time interval}      & Pulsar & Avg. emission      & Aug. 4, 2008 - July 31, 2012 & 	54682 - 56139 \\
\bottomrule
\end{tabular}

\label{tab:ana_sources}
\end{table}

\section{ANALYSIS OF THE \fl DATA}
\label{sec:analysis}

The analysis of the \fl data for the AGNs was performed using the Science Tools v10r0p5\footnote{\texttt{http://fermi.gsfc.nasa.gov/ssc/data/analysis/software/}} and the Instrument Response Functions (IRFs) "P8R2\_SOURCE\_V6" provided by the \Fermi\ collaboration\footnote{\texttt{http://www.slac.stanford.edu/exp/glast/groups/canda/lat\_Performance.htm}}. 

The gamma-ray emission from the 2 blazars was investigated between 70~MeV and 300~GeV (100 MeV to 300 GeV for the Vela pulsar) energies using the \texttt{gtlike} routine to maximise the binned likelihood function \citep{mattox_like}. The data were extracted from a square region $30^{\circ}\times30^{\circ}$ centred on the position given by the 3FGL catalogue using events with evtclass $=128$ and evtype $=3$\footnote{These parameters for the event selection are the suggested ones for most of the Fermi-LAT analysis. This combination selects photon of SOURCE class that left a signal in both the front and the back part of the tracker. More information at {\tt http://fermi.gsfc.nasa.gov/ssc/data/analysis/documentation/Cicerone/Cicerone\_Data/LAT\_DP.html}}.

The source parameters were obtained fitting a model for each Region of Interest (RoI). These models contain the contribution of all the sources within 30 degrees from the centre and thus includes sources outside the RoI. However all the sources more distant than 5 degrees where fixed to the catalogue value. To take into account the diffuse emission we used the \Fermi\ templates \texttt{iso\_P8R2\_SOURCE\_V6\_v06} and \texttt{gll\_iem\_v06} for the isotropic and galactic diffuse emission\footnote{\texttt{http://fermi.gsfc.nasa.gov/ssc/data/access/lat/BackgroundModels.html}} respectively.

To determine the contribution of the background sources we have fitted the RoIs in 2 steps removing all of the sources with test statistic value less than 4 ($\sim 2\sigma$) for the null hypothesis of not having the source in that location. For these flaring sources this procedure was done on longer time intervals to avoid the influence of statistical fluctuations due to very short time-scales. To correctly estimate the flux at energies below 100 MeV and reduce the level of systematic uncertainties on the effective area, the analysis made use of the energy dispersion correction.

After this procedure to fix the background sources, we performed a final fit on the pulsar and the flaring state of the AGNs saving the parameters of the stretched exponential cut-off model and the value of the covariance between them. The spectral points were instead obtained by fitting the central source with a simple power-law and storing the normalisation value in each energy bin. The points of the Spectral Energy Distribution (SED) were then computed for bins with a test statistic (TS) value of at least 9 ($\sim 3\sigma$). The error bars computed with the Science Tools provided by the \fl are the Gaussian approximation of the Poissonian statistic (using the square root of the number of counts). This approach is incorrect for the case of a very small number of photons in the bin. For this reason, if the reconstructed number of photons in the energy bin was less than 10, the error bars were rescaled to take into account the correct $1\,\sigma$ confidence interval for the proper treatment of the Poissonian theory, which can be found in \citet{gehrels_1986}.

\subsection{Vela pulsar}

\begin{figure}
\centering
\includegraphics[scale=0.45]{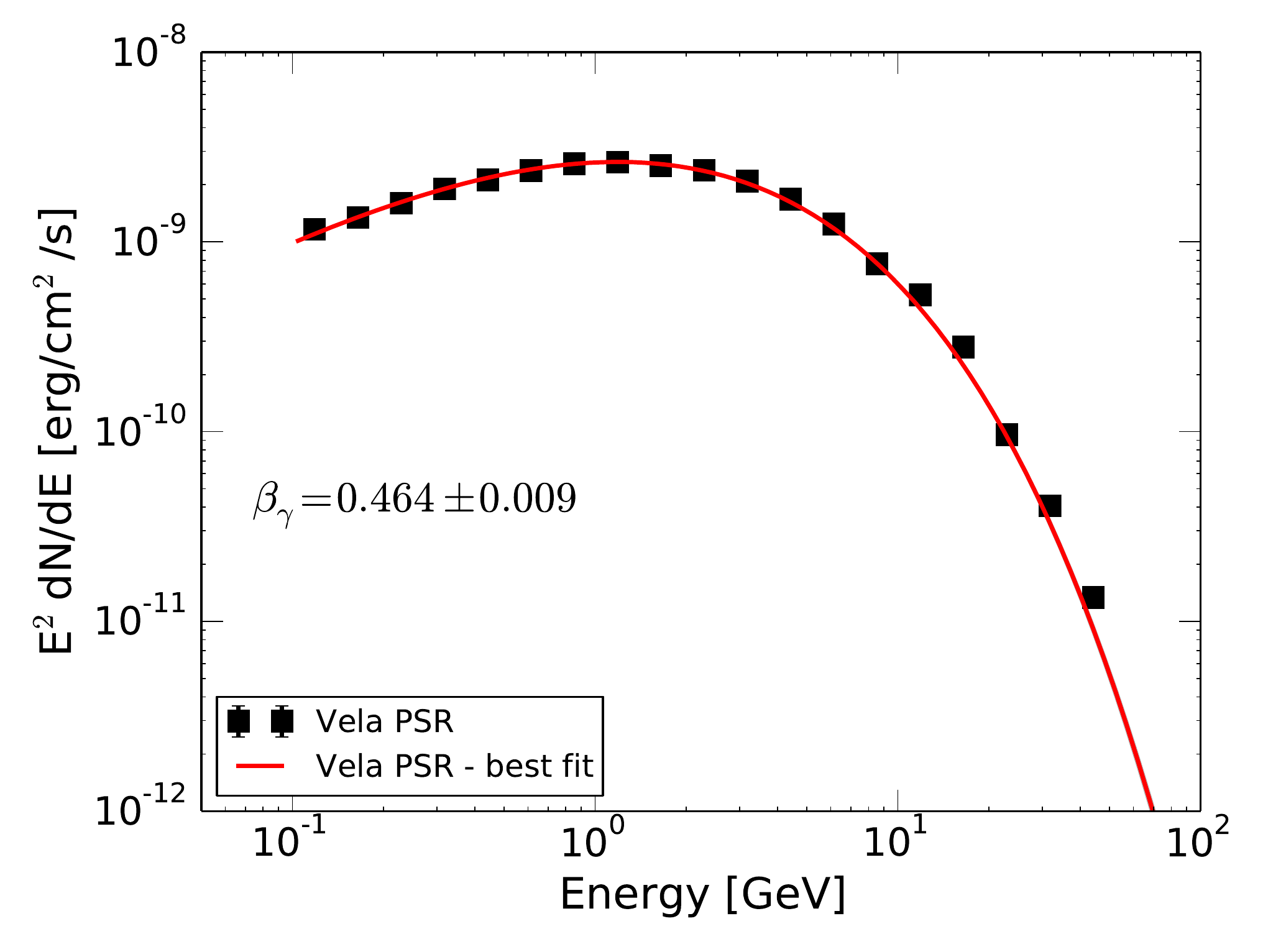}
\caption{SED of the Vela pulsar averaged over 4 years of data. The thick red curve is the best fit model and the shaded area represents the 1 $\sigma$ confidence band (not visible due to the small statistical uncertainties).}
\label{fig:vela_fermised}
\end{figure}

The Vela pulsar is the brightest persistent source in the GeV energy range \citep{vela_fermi2010}. In our analysis we used the averaged emission of the pulsar over 4 years of data using the same time interval of the 3FGL catalog. Due to the large amount of data collected by the \fl for this object, the corresponding spectral parameter constraints were the strongest compared to the other two sources and we can consider this as our best case study. We report the best fit values for all the parameters in table \ref{tab:vela_ph_res_flare} and the SED with the datapoints in Figure \ref{fig:vela_fermised}.

\begin{table}[ht]
\centering
\caption{Fit of the photon spectrum with a power-law with stretched exponential
cut-off for the Vela pulsar as obtained by the \emph{gtlike} routine}
\footnotesize
\begin{tabular}{c|c}
\toprule
Parameter               & Value \\ 
\midrule
\midrule
$N$ {[}ph/cm$^2$/s/GeV{]} & $\left(1.39^{+0.12}_{-0.10}\right)10^{-5}$ \\

$\Gamma$               & $1.019 \pm 0.011$                    \\

$E_c$ {[}GeV{]}        & $0.238\pm 0.016$                     \\

$\beta_{\gamma}$       & $0.464 \pm 0.009$                  \\

$E_s$ (fixed) {[}GeV{]} & 0.83255                       \\
\bottomrule
\end{tabular}
\label{tab:vela_ph_res_flare}
\end{table}

The parameter we are most interested in is the value of the parameter $\beta_{\gamma}$, which distorts the cut-off. For this dataset it was obtained $\beta_{\gamma}=0.464 \pm 0.009$.

From this result we can exclude the possibility of $\beta_{\gamma}=1$, namely a simple exponential cut-off function. A $\beta_{\gamma}$ smaller than one, for the case of pulsars, can be explained as the outcome of a superposition of the spectra during the various phases of the pulse \citep{2pc_fermi2013}. Besides superposition effects, sub-exponential cut-offs can also naturally arise when taking into account the emission in the transition regime between curvature and synchrotron radiation, as shown by \citet{synchro_curv2015}.  With the level of statistics that \fl already has, it would also be possible to study separately the various phases of the pulsed emission. However, a deeper analysis of the Vela spectrum is beyond the scope of this manuscript.

Another aspect that should be noted when dealing with a sub-exponential cut-off, is that, having $\beta_{\gamma} < 1$, the cut-off value shifts to lower energies, with the bend of the spectrum starting at much lower energies. For the Vela pulsar we are already in the cut-off region at energies of 250 MeV having a value for $E_c = 0.238 \pm 0.016$ GeV.

\subsection{3C~454.3 flare}

\begin{figure}[ht]
\centering
\includegraphics[width=0.6\textwidth]{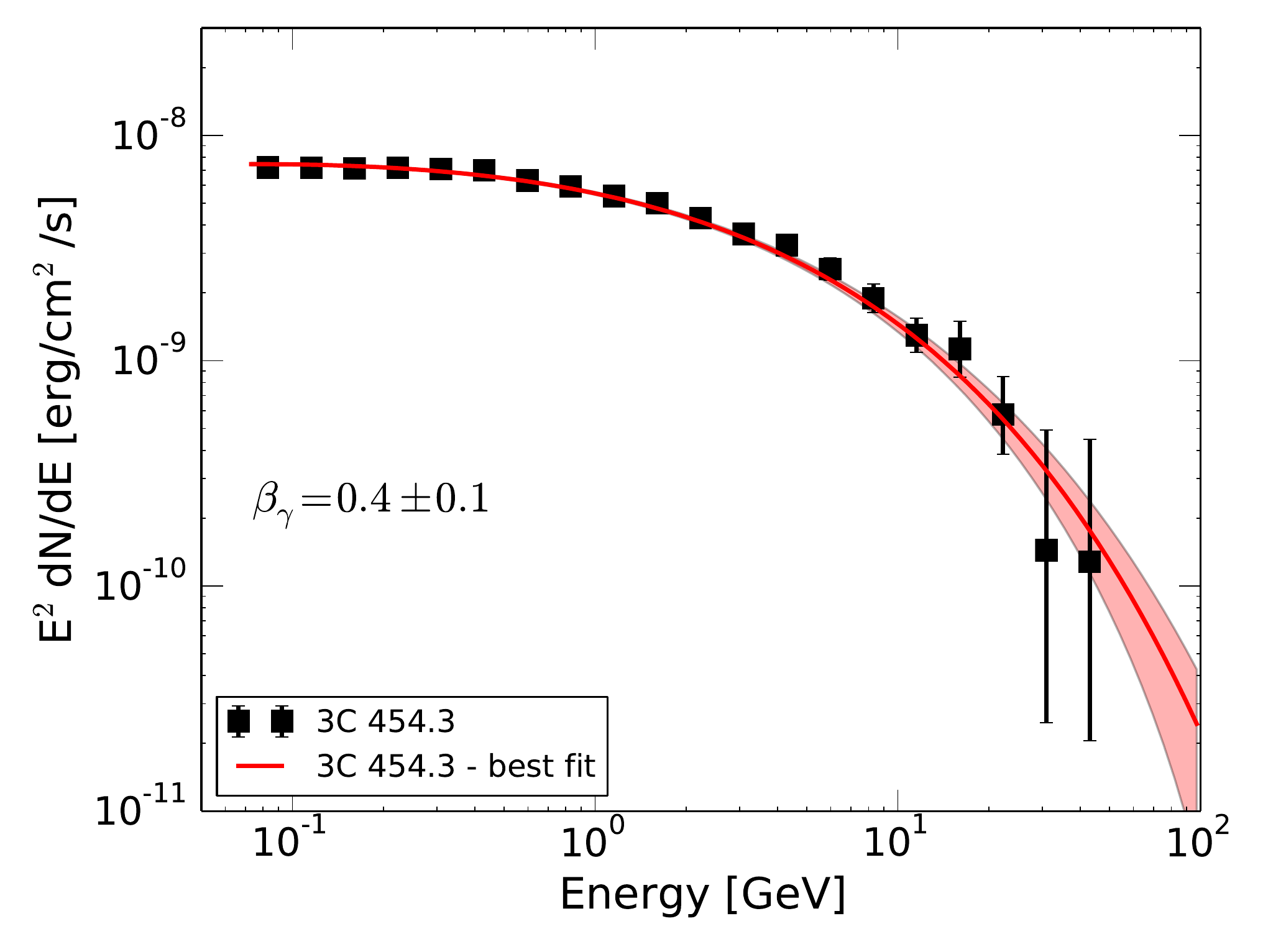}
\caption{SED of the blazar 3C~454.3 during its flaring phase with the value of the parameter beta. The thick red curve is the best fit model and the shaded area represents the 1 $\sigma$ confidence band.}
\label{fig:3c454_fit}
\end{figure}

The FSRQ 3C~454.3 is the brightest AGN in the GeV band observed by \fl. It is a highly variable source located at z=0.859 \citep{3fgl_cat}. 
For our study we analysed the brightest flare detected by the \fl in November 2010 when the source reached an integrated flux above 100 MeV of $\sim 8 \times 10^{-5}$~ph~cm$^{-2}$~s$^{-1}$ \citep{fermi_3c454}. The interval to identify the flaring phase was taken from \citet{fermi_3c454} and analysed with the latest tools provided by the \Fermi\ collaboration. The produced SED is shown in figure~\ref{fig:3c454_fit} where we also show the fit of the power law with stretched exponential cut-off (PLSEC) and the 1 $\sigma$ contour based on statistical uncertainties.

The analysis for 3C~454.3 cannot provide a constraint as strong as the one obtained for the Vela pulsar. In this case $\beta_{\gamma} = 0.4 \pm 0.1$. The results of the value of the other parameters are reported in table~\ref{tab:3c454_ph_res_flare} where we can notice the asymmetry between the lower and upper 1 $\sigma$ interval.

\begin{table}[ht]
\centering
\caption{Fit of the photon spectrum with a power-law with stretched exponential
cut-off for 3C~454.3 as obtained by the \emph{gtlike} routine}
\footnotesize
\begin{tabular}{c|c}
\toprule
Parameter               & Value \\ 
\midrule
\midrule
$N$ {[}ph/cm$^2$/s/GeV{]}  & $\left(4.7^{+3.9}_{-1.2}\right)10^{-5}$     \\
$\Gamma$                & $1.87^{+0.08} _{-0.12}$                       \\
$E_c$ {[}GeV{]}         & $1.1^{+1.6}_{-0.9}$                   \\
$\beta_{\gamma}$        & $0.4 \pm 0.1$                      \\
$E_s$ (fixed) {[}GeV{]} & 0.41275\\
\bottomrule
\end{tabular}
\label{tab:3c454_ph_res_flare}
\end{table} 

As a FSRQ blazar, we expect the gamma-ray peak of this source to be produced by Inverse Compton interactions on external photon fields like disc emission or the Broad Line Region. If we assume that these interactions are happening in the Thomson regime, we obtain a value $\beta_e = 1.3 \pm 0.6$ where the big uncertainty is related to the indirect measurement. Alternatively, an SSC model would require $\beta_e=2.7\pm0.9$ leading to a very steep cut-off.
A different explanation that is also compatible with the values we obtained is the emission via proton synchrotron due to interaction between the jet of the source and a red giant star \citep{2013ApJ...774..113K}. In this case the proton spectrum would have a simple exponential cut-off that via synchrotron emission would produce gamma-rays with $\beta_{\gamma}=1/3$. The importance of having a precise measurement of the $\beta_{\gamma}$ parameter is crucial for characterising the interplay of acceleration and radiative cooling during flaring states. The poor constraint we have presently prevents the motivation for further speculation on the possible origin of this value. 

\subsection{3C~279 flare}

This FSRQ is historically known to be a variable gamma-ray emitter, already detected by EGRET \citep{hartman1992_3c279}. Its redshift is $z = 0.536$ \citep{lynds1965}. This bright AGN underwent a very bright flare in June 2015 \citep{agile_3c279_2015atel, fermi_3c279_2015atel} that showed minute-scale variability \citep{2016ApJ...824L..20A}. Our analysis refers to the 3 days with the highest flux \citep{3c279_2015flare} analysed with the latest analysis tools provided by the \fl Collaboration.

\begin{figure}[ht]
\centering
\includegraphics[width=0.6\textwidth]{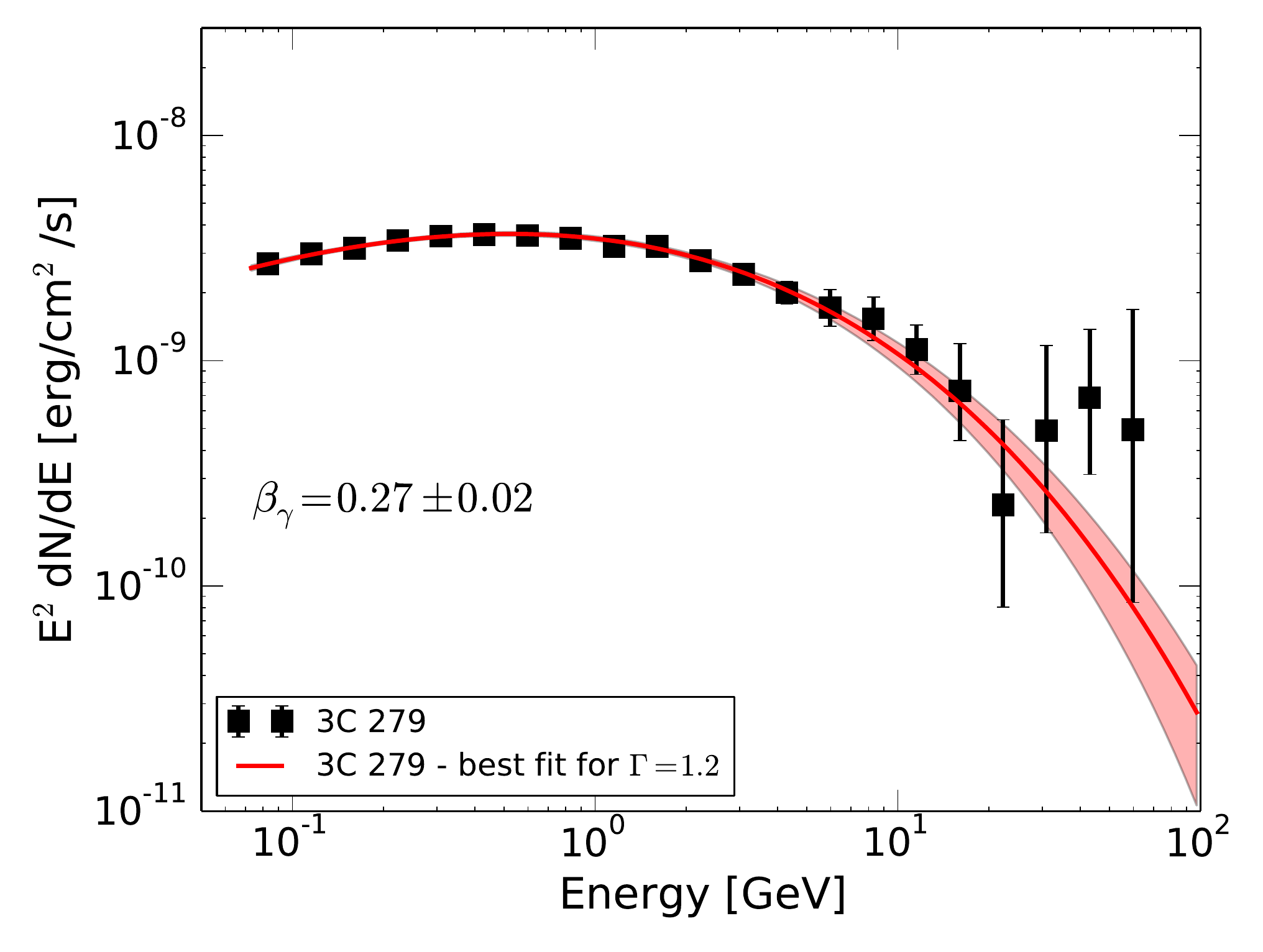}
\caption{SED of the blazar 3C~279 during its flaring phase with the parameter $\beta_{\gamma}$. The thick red curve is the best fit model and the shaded area represents the 1 $\sigma$ confidence band. The parameter value and the confidence band were derived after fixing the photon index $\Gamma$ to 1.2.}
\label{fig:3c279_ph_fit}
\end{figure}

The SED of the source obtained integrating the emission over this time interval is shown in figure~\ref{fig:3c279_ph_fit}.

For this AGN the \emph{gtlike} routine could not converge for the case with all parameters of the modified exponential cut-off model left free. For this reason we performed separate fits fixing the photon index to the values 1.2, 1.4 and 1.6. The choice of these values was motivated by the hard spectrum measured in the X-rays by the \emph{Swift}-XRT instrument. \cite{2016ApJ...824L..20A} report a value of $\Gamma_{X}=1.17 \pm 0.06$ during the peak of the flare, while \cite{2015ATel.7668....1P}, in an ATel report a value of 1.4. A joint fit of the \emph{Fermi}-LAT data and the measured luminosity in the X rays gives instead something close to 1.6. Given these values, we decided to show in the SED in figure~\ref{fig:3c279_ph_fit} only the case for $\Gamma=1.2$ and focus our analysis  on it. The values for the other fits are given in Table 4.

With the photon index parameter fixed, the stretching of the cut-off is constrained to $\beta_{\gamma}=0.27\pm0.02$ where the effect of the missing degree of freedom in the fit reduces considerably the uncertainty on the measurement. One should note that this constraint on the stretching parameter follows the prior assumption on the photon index. With this caveat in mind, we highlight that a value of $\beta_{\gamma}$ close to 0.3-0.4 can be explained either by a simple exponential cut-off in the primary particles external Compton, or by proton synchrotron and $\beta_e \sim 2$ for an SSC scenario.

\begin{table}[ht]
\centering
\caption{Fit of the photon spectrum with a power-law with stretched exponential
cut-off for 3C~279 for the different choices of the photon index. The plot of the SED with the best fit function using $\Gamma=1.2$ can be seen in figure \ref{fig:3c279_ph_fit}}
\footnotesize
\begin{tabular}{c|c|c|c}
\toprule
Parameter                  &     $\Gamma=1.2$              & $\Gamma=1.4$          & $\Gamma=1.6$ \\
\midrule
\midrule
$N$ {[}ph/cm$^2$/s/GeV{]}  & $\left(2.8^{+0.8}_{-0.6}\right)10^{-4}$   & $\left(8.6 \pm 1.0\right)10^{-5}$   & $\left(3.7^{+0.3}_{-0.2}\right)10^{-5}$\\
$\Gamma$  (fixed)       & $1.2$                                     &     $1.4$                          & $ 1.6$ \\
$E_c$ {[}GeV{]}         & $\left( 8.4^{+6.6}_{-4.1} \right)10^{-3}$ &   $0.10 \pm 0.04$                  & $0.81^{+0.16}_{-0.15}$\\
$\beta_{\gamma}$        & $0.27\pm0.02$                             &   $0.34\pm0.03$                    & $0.46 \pm 0.04$\\
$E_s$ (fixed) {[}GeV{]} & \multicolumn{3}{c}{0.341966}\\
\bottomrule
\end{tabular}
\label{tab:3c279_ph_res_flare}
\end{table}

\section{FUTURE POTENTIAL OF GROUND BASED INSTRUMENTS}

The possibility to more sensitively explore the cut-off region of GeV sources may be brought about through an increase of effective area of the gamma-ray instrument. This can be achieved through a lowering of the energy threshold of ground based Cherenkov telescopes. These instruments have already proven themselves to be able to reach a minimum energy close to few tens of GeV under particular conditions as shown by MAGIC \citep{magic_low2012} and HESS\footnote{preliminary result on the detection of the Vela pulsar at 30~GeV. Reported in the news of 27~June~2014 in {\tt https://www.mpi-hd.mpg.de/hfm/HESS/}} collaborations. In this section we demonstrate the improvement possible through such an increase in the effective area at energies around tens of GeV, for our sample of bright \Fermi\ sources.

The idea of pushing the energy threshold of the Imaging Air Cherenkov Telescopes (IACTs) to energies below $\sim 10$~GeV has already been explored in the potential future Cherenkov telescope array, 5@5 \citep{cinque}. In their design, 
the array consisted of 5 big ($\sim 20$~m in diameter) Cherenkov telescopes at an altitude of 5~km above sea level, providing the opportunity to reach down to an energy of 5~GeV.

The construction of the \emph{Cherenkov Telescope Array} (CTA) is planned to start in 2017\footnote{{\tt https://portal.cta-observatory.org/Pages/Preparatory-Phase.aspx}}. This will consist of two sites, one in the northern hemisphere, more optimised to the study of extragalactic objects, and one in the southern hemisphere, enhanced for TeV observations of the galactic plane. Each of them will consist of an ensemble of several Cherenkov telescopes of various diameters to explore different energy bands of the gamma-ray spectrum as described by \citet{CTA-design}.

In this study we focus on the impact of the performance of the CTA observatory on the determination of the spectral parameters of the sources we have studied in the previous section, comparing the SEDs that we obtained with the \fl and the SEDs that we would expect from CTA. For this section we make use of instrument response functions for a preliminary design of the southern array\footnote{{\tt https://portal.cta-observatory.org/Pages/CTA-Performance.aspx} with simulations dated 2015-05-05}, based on the study of \citet{CTA_simulation}. We highlight that these IRFs are still very preliminary as the observatory is not yet in place. The results presented in this section depend on the foreseen performances of the observatory. To make the assumption clearer and to facilitate the understanding, we performed analytical parametrizations of the given IRFs.

Since in our study we are dealing with 2 flaring AGNs, and in general are interested in the possibility of constraining the spectra of bright flaring objects, we have based our studies on the simulations done by the CTA consortium with the optimisation for an observation time of 0.5 and 5 hours to highlight the significant improvement already achievable on short time-scales. For the Vela pulsar we instead simulate only the outcome of a 5 hours observation, which potentially represents the data taken during a single night.


To extract the expected flux in a hypothetical observation from CTA, we used a parametrization of the expected collection area, the background rate and of energy resolution. The collection area was described with a triple smooth broken power-law of the form: 

\begin{equation}
A_{eff}(E)=A\left(\frac{E}{B_1}\right)^{a}\left(1+\frac{E}{B_1}\right)^{b}\left(1+\frac{E}{B_2}\right)^{c} \, \mbox{m}^2
\label{eq:aeff}
\end{equation}
where the parameters $B_1$ and $B_2$ are the positions of the breaks. The maximum difference between this curve and the actual estimate remaining less than 20\%. The background rate after gamma/hadron separation has been approximated instead with a simple power-law of the form:
\begin{equation}
B(E)=N\left(\frac{E}{0.1 \mbox{ TeV}}\right)^{a} \, \mbox{Hz}
\end{equation}

The parametrisation of the background rate can be as far from the simulations as 60\% of the actual estimated rate. However the results are not strongly influenced by the actual level of background due to the extreme brightness of the sources investigated here, with only an increase of several orders of magnitude in the background level being sufficient to lead to noticeable effects to our results. The value of the parameters for these IRFs at 0.5 and 5 hours are reported in tables~\ref{tab:paraeff} and \ref{tab:bkglev}.

\begin{table}
\centering
\textsc{Effective area}\\
\begin{tabular}{ccc}
\toprule
Parameter & 0.5 hours & 5 hours   \\
\midrule
$A$   [$m^2$]& 17461     & 22064       \\
$B_1$  [TeV] & 0.026 & 0.027  \\
$B_2$  [TeV] & 2.86  & 4.65   \\
$a$   & 5.47      & 5.15      \\
$b$   & -4.29     & -4.07   \\
$c$   & -1.23     & -1.18   \\
\bottomrule
\end{tabular}
\caption{Parameters for the parametrization of the effective area for the 0.5 and 5 hours case.}
\label{tab:paraeff}
\end{table}

\begin{table}
\centering
\textsc{Background level}\\
\begin{tabular}{ccc}
\toprule
Parameter & 0.5 hours & 5 hours   \\
\midrule
$N$ [Hz] & 0.0255 & 0.0279\\
$a$ & -1.717 & -1.857\\
\bottomrule
\end{tabular}
\caption{Parameters for the parametrization of the background level after cuts for the 0.5 and 5 hours case.}
\label{tab:bkglev}
\end{table}

The energy resolution was instead modelled using a smooth broken power law:
\begin{equation}
\frac{\Delta E}{E} = 0.0468 \left( \frac{E}{0.64 \mbox{ TeV}} \right)^{-0.59}\left( 1+\frac{E}{0.64 \mbox{ TeV}} \right)^{0.69}
\end{equation}
with an accuracy toward the simulations of less than 8\%. The energy bias was extracted from the migration matrix available together with the other IRFs of the southern site and approximated with the exponential of a power-law function:
\begin{equation}
\frac{E_R-E_T}{E_T}=\exp\left[ -\left( \frac{E_T}{0.023 \mbox{ TeV}} \right)^{2.43} \right]
\end{equation}
where $E_T$ is the true energy of the event and $E_R$ is the reconstructed one. This function was able to approximate well the energy bias near threshold giving a value of the bias of 0.5 at 20 GeV that drops quickly and becomes negligible already around 40 GeV.

From the parameters in table~\ref{tab:paraeff} one can already appreciate the potential of the instrument with respect to the \Fermi\ satellite. The case in favour of ground based telescopes is in the much larger effective area: while \Fermi\ can count only $\sim1$ m$^2$ across the energy range, due to the Cherenkov light pool having a radius of $\sim100$~m, a ground based telescope can in principle detect $\sim 10^4$ more photons in the same time interval for energies below 100 GeV.

Through the convolution of the effective area with the source flux, we computed the expected count rate at the CTA detector. With these count rates, applying Poissonian inference, it was possible to simulate spectral points that CTA would be able to recover for our sources after an observation time of 0.5 and 5 hours. To do this we extrapolated the flux level starting from the \fl best fits, convolved the flux from these fits with the effective area of the CTA observatory, applying the EBL absorption for the 2 extragalactic sources. The EBL model used was the one developed by \citet{Franceschini_EBL2008}. At this point the expected events were drawn from a probability density function that matched the shape of the expected differential count rate at the detector with the total number of events randomly taken from a Poissonian distribution according to the expected total number of photons.

To properly simulate the response of CTA, we smeared the \emph{true} distribution with the parametrized IRFs and successively unfolded this \emph{measured} sample of events to recover the \emph{reconstructed} datapoints. The unfolding of the measured distribution of photons was performed via the \texttt{RooUnfold} package\footnote{\texttt{http://hepunx.rl.ac.uk/~adye/software/unfold/RooUnfold.html}} \citep{Adye:2011gm}. The unfolding procedure used an iterative Bayesian approach trained on a large test dataset to recover the response matrix with arbitrary bin size. To correctly take into account the background level, the distribution injected in the unfolding routine was the sum of the signals of the source and background. The uncertainties on each bin take into account the covariance matrix, which is particularly relevant near the threshold. With this operation we were able to derive the number of expected counts in each of the 10 logarithmically spaced new energy bins spanning from 20 GeV to 2~TeV (11 bins for the 5 hours case).

The actual source counts and the 1~$\sigma$ errorbars were instead derived starting from the total number of counts in the bin $N=S+B$, where $S$ are the counts coming from the source and $B$ are the background counts, known with good accuracy for the observation.

The lower and upper limits on the source counts were then extracted following a ``classical approach'' \citep{kraft_low1991}:
\begin{equation}
S_{up}=N_{up}-B
\end{equation}
and
\begin{equation}
S_{low}=N_{low}-B.
\end{equation}
We avoid the case of $B>S$ in our calculation, defining only high and low limits for a 68\% confidence interval.

\begin{figure}[ht]
\centering
\includegraphics[scale=0.3]{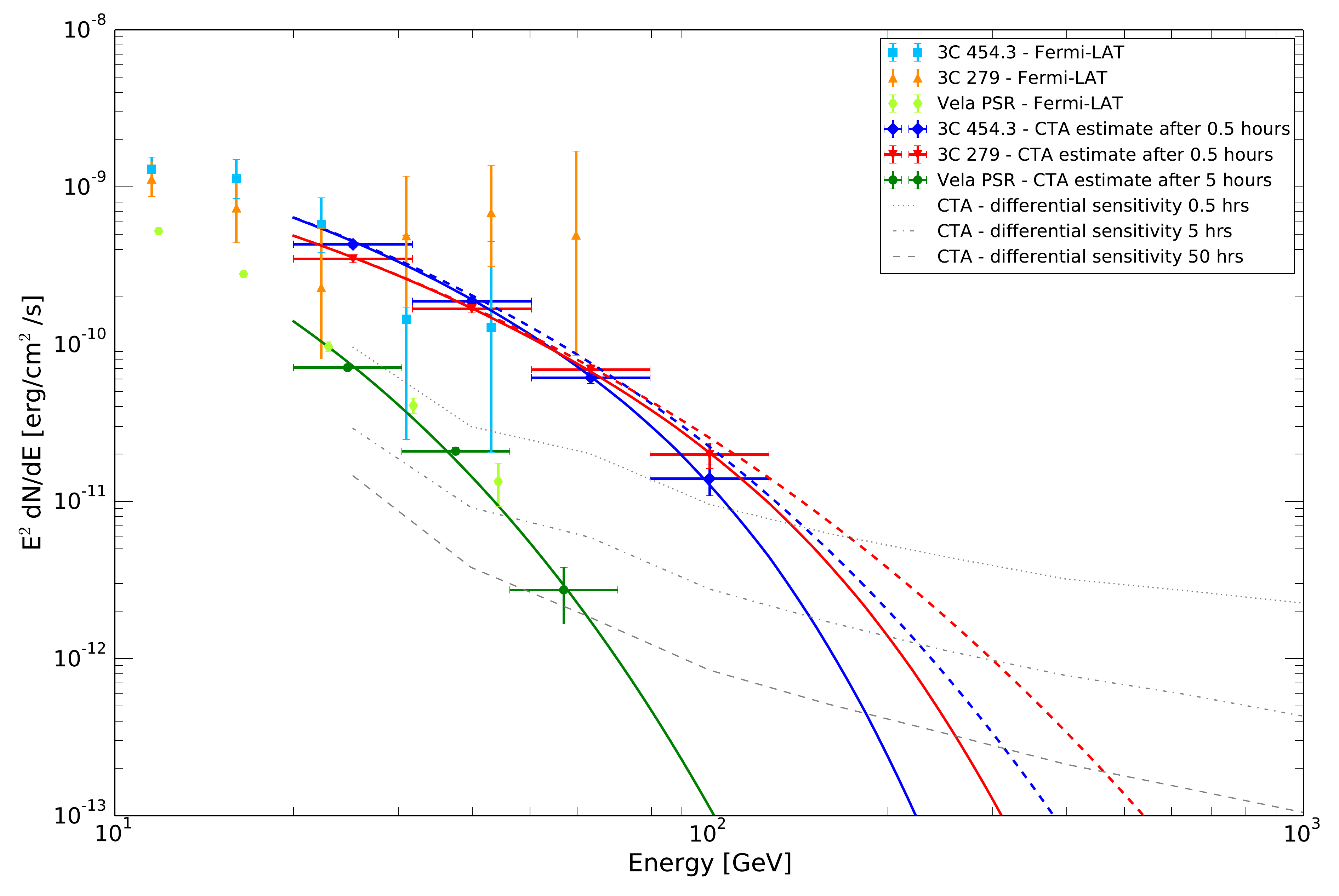}
\caption{SED above 10 GeV with the \fl points and the CTA estimate after 30 minutes of observation for 3C~454.3 and 3C~279 and after 5 hours for the Vela pulsar (blue, red and green respectively). The solid lines are the observable extrapolated spectra that in the case of the AGNs have been absorbed according to the EBL model of \citet{Franceschini_EBL2008}, while the dashed lines represent the de-absorbed extrapolation. The grey lines are the differential sensitivity of CTA-South after 0.5, 5 and 50 h of observation.}
\label{fig:sed_flcta}
\end{figure}

The resulting SED showing both the \fl points above 10~GeV and the CTA ones is shown in figure \ref{fig:sed_flcta}. The \fl points are  taken from the previous analysis described in Sect.~\ref{sec:analysis}, obtained with an integration time of a few days. As can be seen in the plot, for the 2 bright AGNs considered, up to 100~GeV the CTA points are above the 5$\,\sigma$ detection limit: flares with brightness similar to the ones analysed here will therefore be easily detected. For the Vela pulsar we are instead deeply in the cut-off region and on such short time there is not enough statistics to reconstruct points above 80 GeV, nevertheless, the size of the reconstructed error bars compare very well with what Fermi was able to achieve with 4 years of data.

We next assess the improvement in our ability to constrain the parameter $\beta_{\gamma}$ for the 2 blazars by fitting the combined \Fermi\ and CTA datasets shown in figure \ref{fig:sed_flcta}. For the Vela pulsar the addition of the 3 CTA datapoints is not able to improve significantly the constraints, due to the high level of statistics that the \fl achieved with an integration of 4 years. The $\chi^2$ minimization was performed through a Markov-chain Montecarlo (MCMC) using the PYTHON tool  \emph{emcee}\footnote{\texttt{http://dan.iel.fm/emcee/current/}} developed by \citet{emcee_paper}, based on the technique of the ensemble samplers with affine invariance \citep{mcmc_ens_aff}.

\begin{figure}
\centering
\subfloat[3C~454.3]{\includegraphics[width=0.45\textwidth]{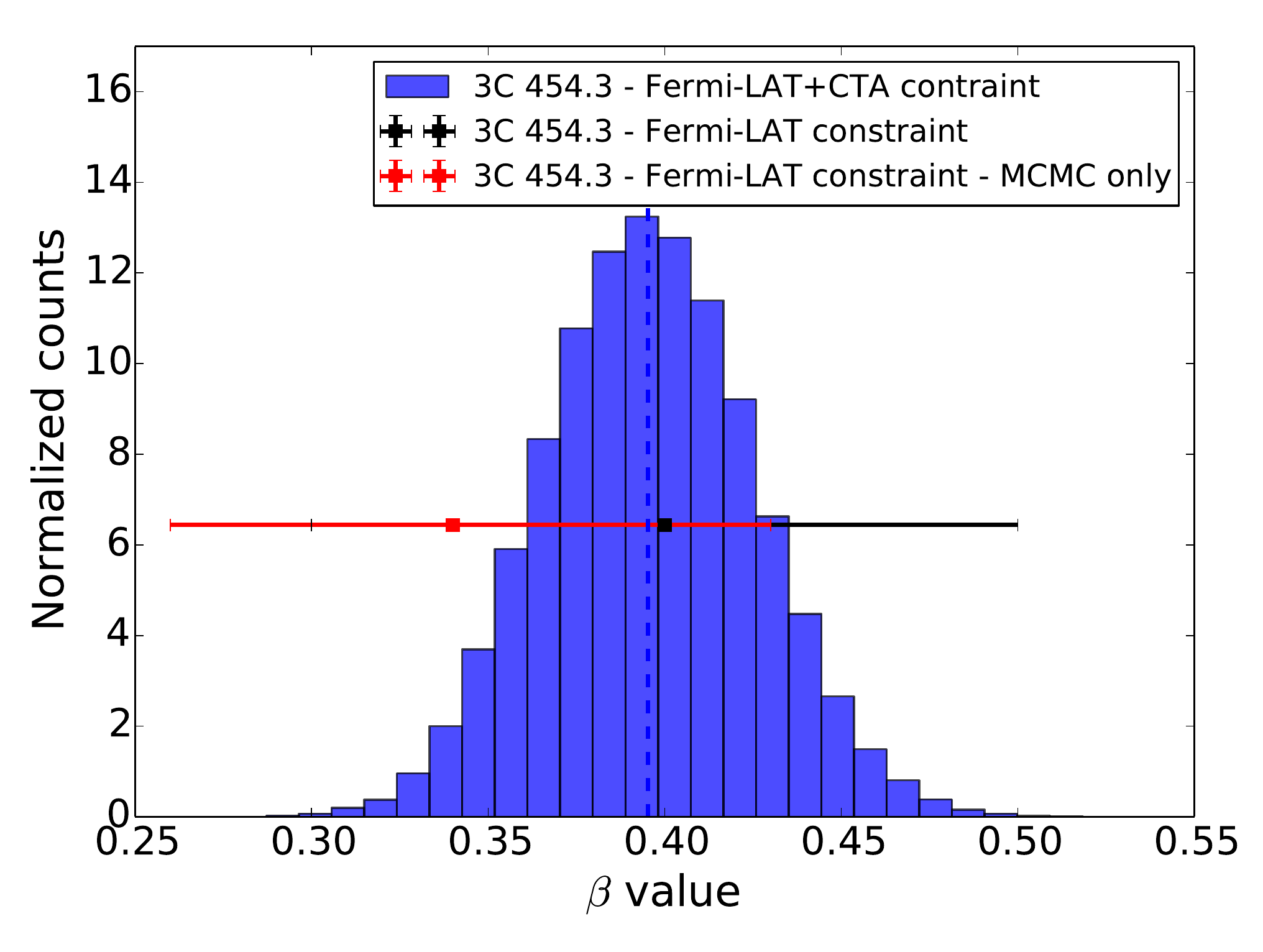}\label{fig:3c454_betaflcta}}
\subfloat[3C~279]{\includegraphics[width=0.45\textwidth]{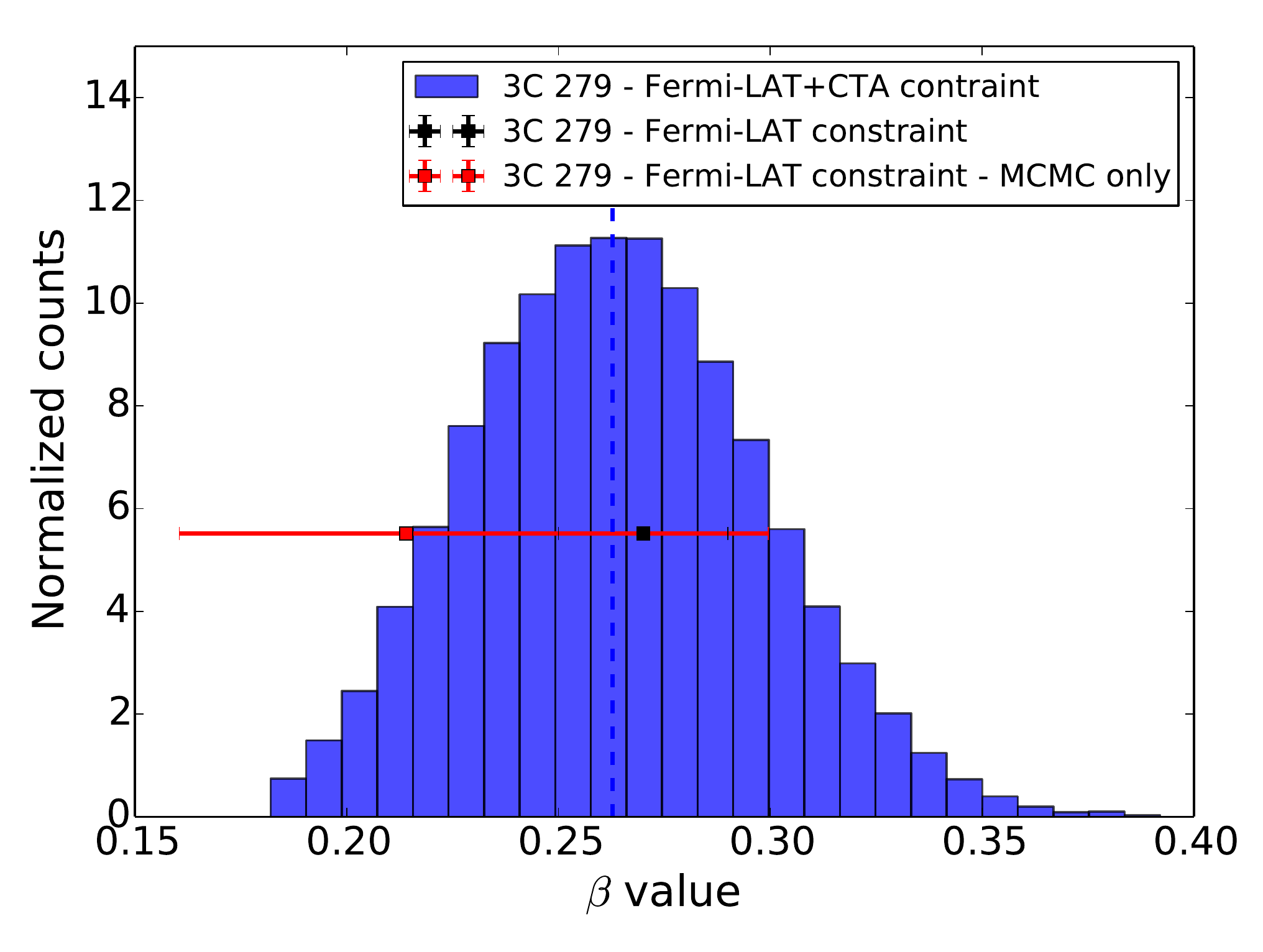}\label{fig:3c279_betaflcta}}
\caption{$\beta_{\gamma}$ posterior distribution of the joint dataset. In panel \ref{fig:3c454_betaflcta} there is the $\beta_{\gamma}$ distribution for the blazar 3C~454.3 while in panel \ref{fig:3c279_betaflcta} we report the one for the blazar 3C~279. The full histogram represents the distribution of $\beta_{\gamma}$ for the joint fit while the black and red bars correspond to the 1 sigma confidence interval obtained when fitting the \fl points only using the official tools and the MCMC method, respectively. Note that in the fit of the joint 3C~279 dataset, the photon index was left free to vary with a consequent shift in the value of the parameters. The width of the posterior distribution in this case remains comparable to the size of the 1 $\sigma$ confidence interval obtained with the \fl only data and with a smaller number of degrees of freedom.}
\label{fig:beta_flcta}
\end{figure}

The starting point for the MCMC routine was the best fit model from the \fl data and letting 100 parallel walkers to run for 24000 steps with a burn-in of 100 steps. For the fitting procedure the $\chi^2$ contribution for the CTA dataset utilised the covariance matrix obtained from our unfolding analysis. Prior to this fitting the CTA data are deabsorbed on the EBL to obtain the intrinsic spectrum at source. The resulting constraints on the $\beta_{\gamma}$ parameter for the 2 bright blazar flares are reported in figure \ref{fig:beta_flcta}, along with a comparison with the result from the posterior distribution obtained with the \fl data alone and with the result obtained with the official \fl tools. The mean value and the RMS of these histograms are reported in table~\ref{tab:tab_jointfit} for the case of 0.5 hours and, using the same methodology, for 5 hours of observation. The difference between the values of $\beta_{\gamma}$ for 3C~279 depends on the fact that an MCMC fit on the \fl data only converges to parameters values that are different from the ones used to extract the CTA data points as visible in figure~\ref{fig:3c279_betaflcta}.

This result clearly demonstrates that we are able to reduce the uncertainty on the beta parameter down to the $\sim10\%$ level by adding the data that CTA could collect in just 0.5 hour.

\begin{table}
\centering
\caption{Value of mean and RMS of the $\beta_{\gamma}$ parameter after the fit of the \fl and CTA estimated data for observation time of 0.5 hours and 5 hours.}
\footnotesize
\begin{tabular}{c|c|c}
\toprule
Object         & $\beta_{\gamma}$ after 0.5 hrs (ratio error/value) & $\beta_{\gamma}$ after 5 hrs (ratio error/value)     \\
\midrule
{\bf 3C~454.3} & $0.40 \pm 0.03 $ (0.08)   & $0.40 \pm 0.02$ (0.05) \\
{\bf 3C~279}   & $0.26 \pm 0.03 $ (0.12)   & $0.26 \pm 0.02$ (0.08) \\
\bottomrule
\end{tabular}
\label{tab:tab_jointfit}
\end{table}

\section{CONCLUSIONS}

In this paper we investigated the spectra of a sample of some of the brightest sources observed by \fl, namely the Vela pulsar and the 2 bright FSRQs 3C~454.3 and 3C~279. The shape of the cut-off in the spectra for each of these objects were investigated in order to determine how well constrained they are by the current data set. The sources were instead analysed with the most recent \fl software to improve the level of the statistics. The value of the stretching parameter $\beta_{\gamma}$ retrieved from the analysis of the Vela pulsar is very well defined thanks to the very good statistics obtained through the long exposure. The 2 blazars still suffered from fewer counts above 10~GeV where the photon flux is too low for the \fl to obtain good constraints. The results place mild limits on the primary particle distribution due to the uncertainties from the fits. The level of uncertainty on the parent (electron) cut-off index $\beta_e$, however, is inferred to sit between 30 and 50 percent.

The aim of our study here was to also demonstrate the significant potential that bright GeV objects possess with regards to studies of their underlying particle spectra, and the limitation current instruments place on these studies. While for steady sources the continuous observation by the \fl can give high quality data, for flaring objects the case is different. The limiting factor here being a lack of sufficient statistics in case of transient sources for energies above 10~GeV. Indeed, although the \fl has contributed in a decisive way in our knowledge of the high energy range, it is intrinsically limited by its small effective area that is around 1~m$^2$.

Finally, the benefits promised by an IACT system with a threshold in this energy range for our specific bright object set were investigated. As a reference, we used simulations for the near future of the field: the Cherenkov Telescope Array (CTA) that will be operative within the next few years. We applied simulations for its effective area to the real data coming from our set of bright GeV sources. In particular, for the bright objects we analysed, the statistical uncertainty after 0.5 hours of observation will already be small enough in order to put strong constraints on the spectral shape, with a level of statistic comparable to that of \fl after years of data taking. The role of the \fl is however crucial for two distinct reasons: 1) to give the trigger for bright GeV flares; 2) to extend the spectrum to sub-GeV energies and give constraints at the beginning of the cut-off region, a role that CTA would not be able to take for many extragalactic objects.

Using the data from both instruments we have shown that the cut-off could be defined down to even 10\% precision level, allowing a potential revolution in the understanding of transient objects at high energies, with the chance of capturing an evolution of the cut-off during the flare, while the source balances acceleration and cooling of the primary particles.

The revolution of lowering the threshold of ground based Cherenkov observatories to $\leq 10$ GeV, could express itself in the observation of phenomena never thought before, in the same way \fl was able to observe Crab flares at $\sim$GeV energies. Phenomena like these are still to be completely explained and Nature can still surprise us.

\section*{Acknowledgements}
The authors would like to thank the journal referees that helped improving the quality of the manuscript. This paper has also gone through internal review by the CTA Consortium.

\section*{References}
\bibliographystyle{elsarticle-harv}
\bibliography{mn-jour,bib_rep_br_sources}
\end{document}